\documentclass[aps,amsfonts,prl,twocolumn,showpacs]{revtex4-1}
\usepackage{epsfig,amsmath,amssymb,bm,epsf,graphics,psfrag,verbatim,subfigure}

\def\beq{\begin{equation}}
\def\eeq{\end{equation}}
\def\bea{\begin{eqnarray}}
\def\eea{\end{eqnarray}}

\begin{document}

\title{Frustration and glassiness in spin models with cavity-mediated interactions}
\author{Sarang Gopalakrishnan$^1$, Benjamin L. Lev$^{1,2}$, and Paul M. Goldbart$^3$}
\affiliation{$^1$Department of Physics, University of Illinois at Urbana-Champaign, 1110 West Green Street, Urbana, Illinois 61801 \\ $^2$Departments of Applied Physics and Physics, and E. L. Ginzton Laboratory, Stanford University, Stanford, California 94305 \\ $^3$School of Physics, Georgia Institute of Technology, 837 State Street, Atlanta, Georgia 30332}

\date{August 5, 2011}

\begin{abstract}
We show that the effective spin-spin interaction between three-level atoms confined in a multimode optical cavity is long-ranged and sign-changing, like the RKKY interaction; therefore, ensembles of such atoms subject to frozen-in \textit{positional} randomness can realize spin systems having disordered and frustrated \textit{interactions}. We argue that, whenever the atoms couple to sufficiently many cavity modes, the cavity-mediated interactions give rise to a spin glass. In addition, we show that the quantum dynamics of cavity-confined spin systems is that of a Bose-Hubbard model with strongly disordered hopping but no on-site disorder; this model exhibits a random-singlet glass phase, absent in conventional optical-lattice realizations. We briefly discuss experimental signatures of the realizable phases. 
\end{abstract}

\maketitle

Realizing models of magnetic phenomena and exploring their phases has been a central objective in ultracold atomic physics since the advent of optical lattices~\cite{bloch:review}. Such models (e.g., the Hubbard and Heisenberg models) have been of long-standing theoretical interest as they are believed to offer minimal descriptions of strongly correlated materials~\cite{dagotto}. Unlike real materials, ultracold atomic systems offer the prospect of realizing the theoretical models \textit{exactly}; because many of the models are not solvable, it is hoped that ultracold-atomic realizations will shed light on their properties. 
The central effort to realize magnetism, to date, has focused on the fermionic Hubbard model~\cite{moritz}; however, its magnetic ordering temperature is too low to be readily achievable in current experiments. These difficulties have stimulated an interest in alternative paths to quantum magnetism~\cite{sengstock, ketterle, gorshkov}, of which the present work is an example. 

\begin{figure}
	\centering
		\includegraphics{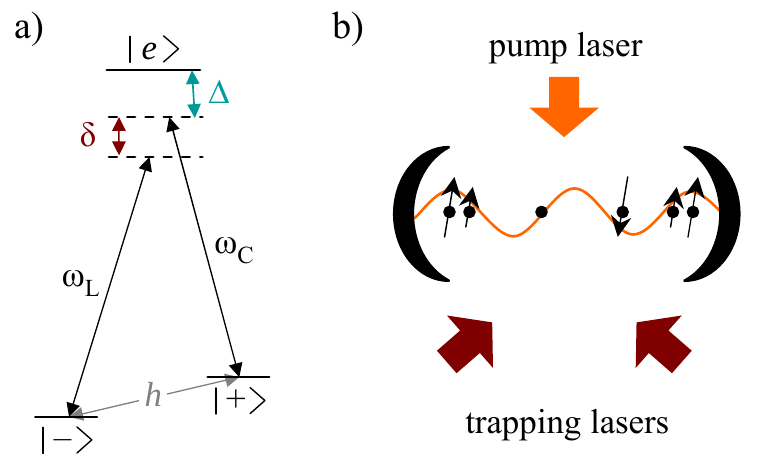}
	\caption{(a)~Level structure of three-level $\Lambda$ atoms, dressed by a pump laser at frequency $\omega_L$, cavity mode(s) at frequency $\omega_C$, and a microwave field represented by $h$. The detuning from two-photon resonance, $\delta$, is assumed to be much smaller than the detuning of laser and cavity photons from the atomic transition, $\Delta$. (b)~Proposed experimental setup. Atoms are tightly trapped by trapping lasers, which are far detuned from the atomic transition, and pumped transversely. Spins are self-organized as discussed in the text for a single-mode cavity, with a sinusoidal mode function as depicted: spins at even antinodes interact ferromagnetically with spins at other even antinodes, but antiferromagnetically with spins at odd antinodes. Spin-spin interactions are strongest for spins trapped at antinodes; therefore, ordering is strongest at antinodes and weakest at nodes.}
	\label{fig:fig1t}
\end{figure}

We introduce a scheme for realizing magnetism---involving $\Lambda$-type three-level atoms [see Fig.~\ref{fig:fig1t}(a)] trapped in a multimode optical cavity---that differs from previous schemes in an essential respect, viz. the range and structure of interactions. Whereas previous schemes have involved contact or dipolar interactions~\cite{gorshkov}, the spin-spin interactions in our scheme, being mediated by cavity modes, are both \textit{long-ranged} (indeed, infinite-ranged for a single-mode cavity) and \textit{oscillatory} in sign. In these respects, they resemble the Ruderman-Kittel-Kasuya-Yosida (RKKY) interaction~\cite{rk}, which underlies, e.g., the physics of heavy-fermion materials~\cite{sigrist} and metallic spin glasses~\cite{binder, weissman}. The present work is concerned chiefly with the latter class of systems, and, in particular, with the fact that long-range, sign-changing interactions between spins facilitate the realization of various frustrated and bond-disordered models. (Analogous realizations have also been proposed, e.g., in photonic band-gap systems~\cite{john:quang} and Coulomb crystals~\cite{bermudez}.) 

The elements of our scheme are $\Lambda$-type atoms (i.e., atoms with the level structure shown in Fig.~\ref{fig:fig1t}) dressed by a configuration of laser and microwave fields suggested in Ref.~\cite{sorensen:prl}; the atoms are assumed to be tightly confined near fixed, random positions inside the cavity. (Alternatively, the spins might arise because of nitrogen-vacancy centers in diamond~\cite{childress}, distributed randomly inside a multimode cavity.) We show that the effective spin Hamiltonian is a variant of that studied in Refs.~\cite{vanHemmen, amit, ioffeJETP, weissman}. By adapting the results of Refs.~\cite{amit, weissman}, we show that, depending on the number of spins per strongly-coupled cavity mode, the low-temperature phase is either a spin glass or a superradiant phase~\cite{ritschprl, vuletic:prl, baumann} (analogous to a cavity-mediated crystal~\cite{us, us:pra}); we discuss how these phases can be distinguished experimentally. In contrast with condensed-matter realizations, the systems considered here allow one to access both regimes in the same system, by changing the mirror spacing and thereby tuning the number of active cavity modes~\cite{siegman}. We note, moreover, that for \textit{quantum} spins, the effective Hamiltonian can be mapped onto a Bose-Hubbard model possessing strictly off-diagonal disorder~\cite{altman:mottglass}. Unlike the diagonally disordered Bose-Hubbard model~\cite{fisher89, mattP}, the off-diagonally disordered version exhibits multiple distinct insulating phases, including a Mott glass phase and a random-singlet glass phase~\cite{altman:mottglass, bhatt:lee, dfisher:singlets}, neither of which has been experimentally observed so far.

\textit{Model.}
We consider $\Lambda$-type atoms whose lower levels (which will be our two spin states, $\vert + \rangle$ and $\vert - \rangle$) are separated by a microwave transition whereas the excited level, $\vert e \rangle$, is separated from both by an optical transition. The $\vert \pm \rangle$ states are assumed to be tightly confined at the intensity extrema of trapping lasers that are far detuned from the $|\pm \rangle \rightarrow | e \rangle$ transition; i.e., the atomic positional degrees of freedom are assumed to be frozen out. (Thus, the physics considered here differs from that of \textit{mobile} spinful atoms~\cite{blacketal, zheng, zoubi09}.) Disorder can be introduced using diffusers (see, e.g., Ref.~\cite{mattP}). The atoms are confined in an optical cavity having multiple degenerate modes, at a frequency red-detuned from the $\vert + \rangle \rightarrow | e \rangle$ transition by $\Delta \sim 1$~GHz; other modes are typically farther-detuned (e.g., by $\sim 15$~GHz for a $1$~cm cavity). Additionally, the atoms interact with a pump laser oriented \textit{transverse} to the cavity axis, red-detuned from the $| - \rangle \rightarrow | e \rangle$ transition by $\Delta + \delta$, where $\delta \simeq 10$~MHz is the detuning from two-photon resonance. (Note that the aforementioned setup generalizes to systems possessing manifolds of ground and excited atomic states, rather than two ground states and one excited state, provided that---as in dysprosium~\cite{LevDy}---the two manifolds have similar $g$-factors.) The microwave $|+\rangle \leftrightarrow |-\rangle$ transition is driven at a weak Rabi frequency that, as we shall see, acts as an effective magnetic field.

Under these conditions, the spin-spin interactions can be understood as follows: an atom in the $\vert - \rangle$ state can scatter a laser photon into a cavity mode, thus changing its state to $\vert + \rangle$; this virtual cavity photon, being $\delta$ higher in energy than laser photons, is reabsorbed into the laser after a time $\sim 1/\delta$.  The reabsorption involves flipping the state of a $\vert + \rangle$ atom (typically a different one from the initial atom) to $\vert - \rangle$. This entire process generates an effective interaction of the form $M(\mathbf{x}_i, \mathbf{x}_j) \sigma_+^i \sigma_-^j / (\Delta^2 \delta)$ between two atoms, where $\mathbf{x}_i$ is the position of the $i^{\mathrm{th}}$ atom and $M$ is a matrix element (derived below) depending on the cavity mode(s). 

We further assume that the cavity photon leakage rate per mode, $\kappa \ll \delta$, and also that the atomic-excited-state decay rate, $\gamma \ll \Delta$. In this ``\textit{dispersive}'' regime, the conservative virtual-excitation processes fall off as $1/\delta$ and $1/\Delta$ respectively, whereas the dissipative processes fall off as $\kappa/\delta^2$ and $\gamma/\Delta^2$ respectively. As argued in Ref.~\cite{us:pra}, the effect of dissipation in this regime can be understood in terms of heating, and need not be explicitly included in the Hamiltonian. Generally, dissipation does not change the mean-field properties even beyond this regime~\cite{domokos11}; we shall revisit this issue in future work. (Note that a weak microwave field must be applied to prevent the spin population from being pumped entirely into the $|+\rangle$ state; however, this field is comparable in strength to the decay processes, and thus much weaker, for small $\kappa/\delta$, than the interaction terms.) Thus, we neglect, in this work, issues such as the nonequilibrium growth of entanglement~\cite{sorensen:prl}.

Hence, upon \textit{adiabatic elimination}~\cite{ritschpra, us:pra} of the state $\vert e \rangle$, the Hamiltonian $\mathcal{H}$ of the atom-light system takes the form

\beq\label{eq:atomlight}
\mathcal{H} = H_{\mathrm{at}} + \sum\nolimits_\alpha \omega_\alpha a^\dagger_\alpha a_\alpha + \frac{\Omega}{\Delta} \sum_{\alpha, i = 1}^N g_\alpha(\mathbf{x}_i) \sigma_-^i a^\dagger_\alpha + \mathrm{h.c.},
\eeq
where $\omega_\alpha$ is the frequency of cavity mode $\alpha$; $a_\alpha$ destroys a cavity photon; $\Omega$ is the strength (i.e., Rabi frequency) of the pump laser; $g_\alpha(\mathrm{x}_i)$ describes the coupling to mode $\alpha$ at the position $\mathbf{x}_i$ of atom $i$; and the $\sigma$ operators are Pauli matrices acting on the atomic ground-state manifold. One can rewrite the coupling $g_\alpha(\mathrm{x}_i)$ as $g \Xi_\alpha(\mathbf{x}_i)$, where $g$ is an overall coupling strength (assumed to be the same for all strongly-coupled modes) and $\Xi_\alpha(\mathbf{x}_i)$ a normalized mode profile. The terms in $H_{\mathrm{at}} = \sum\nolimits_i (h_x \sigma^i_x + h_z \sigma^i_z)$ represent transitions that do not involve the cavity, and are due to the $|+\rangle \leftrightarrow |-\rangle$ microwave driving: $h_x$ is the microwave Rabi frequency, $h_z$ is the detuning, and $\sigma^i$ are the Pauli matrices for atom $i$. In what follows we refer to these terms as ``fields.'' Note that the model described above, while similar in some ways to the multimode Dicke model~\cite{mmdicke}, differs from it in the crucial respect that, in the present case, the different modes have \textit{distinct spatial profiles}; it is this feature, not present in the multimode Dicke model, that enables frustration to be realized.

We now proceed to eliminate the cavity modes perturbatively, thus arriving at an effective model for the spins, valid on timescales $\agt 2\pi/\delta$: 

\beq\label{eq:1mode}
H = H_{\mathrm{at}} + \sum_{\alpha; i < j} \frac{|\Omega(\mathbf{x}_i)|^2}{\Delta^2} \frac{g_\alpha(\mathbf{x}_i) g^*_\alpha(\mathbf{x}_j)}{\delta} \sigma_+^i \sigma_-^j + \mathrm{h.c.}
\eeq
(Note that this result can also be derived from nonequilibrium field theory, as in the spinless case~\cite{us:pra}.)

\textit{Single-mode case.} In what follows we denote the effective spin-spin coupling as $\zeta \equiv |g\Omega|^2 / (\Delta^2 \delta)$. Thus, e.g., for atoms in a single-mode cavity for which $\Xi_\alpha(x) \sim \cos(kx)$, the zero-field Hamiltonian is

\beq
H_{\mathrm{1-mode}} = \frac{1}{2}\zeta \left (\sum\nolimits_i \cos(k x_i) \sigma_+^i \right)\!\! \left( \sum\nolimits_j \cos(k x_j) \sigma_-^j \right) + \mathrm{h.c.} 
\eeq
Because the interaction term can be rewritten as $\mathbf{\sigma}_i \mathbf{\cdot \sigma}_j \equiv \sigma_x^i \sigma_x^j + \sigma_y^i \sigma_y^j$, the system possesses an $O(2)$ symmetry (as the repumping field is negligible). The cavity-mediated interaction is \textit{ferromagnetic} for atoms $\lambda$ apart, but \textit{antiferromagnetic} for atoms $\lambda/2$ apart; therefore, the low-temperature ordered state involves all spins at even antinodes aligned along some direction $\theta$ on the equator of the Bloch sphere defined by $|\pm \rangle$, and all atoms at the odd antinodes aligned along $\theta + \pi$. The interactions, though disordered (as their \textit{magnitude} is position-dependent), are not frustrated in this case. Note that spin-ordering leads to a macroscopic photon population in the cavity mode, as in the self-organization of an atomic cloud~\cite{ritschprl, ritschpra, us, us:pra}---put differently, magnetism is a self-organization of atomic spins rather than positions. (This can be seen, e.g., by replacing the $\sigma$ operators in Eq.~(\ref{eq:atomlight}) by their expectation values.) In the driven, dissipative system (with $h_x, \kappa \neq 0$) such macroscopic occupation corresponds to superradiance. 

\begin{figure}
	\centering
		\includegraphics{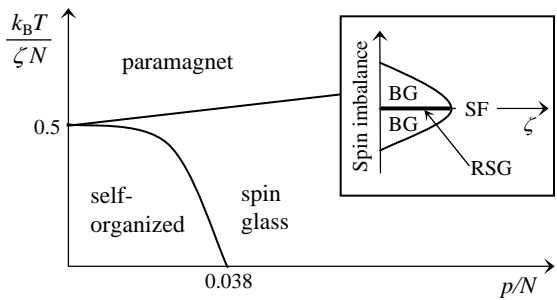}
	\caption{Phase diagram of frustrated spin systems in cavities, as a function of the temperature (vertical axis) and ratio of number of modes, $p$, to number of atoms, $N$. The thresholds for self-organization as $T \rightarrow 0$ and as $p/N \rightarrow 0$ were computed in Ref.~\cite{cook}, and might have geometry-dependent corrections; the boundaries connecting them are schematic. Inset: schematic quantum phase diagram for an off-diagonally disordered XY model, as a function of hopping (i.e., cavity-mediated interaction strength) and spin imbalance, showing SF (``superfluid,'' i.e., magnetically ordered), BG (Bose glass), and RSG (random-singlet glass) phases discussed in the text.}
	\label{fig:phasediag}
\end{figure}

\textit{Multimode case.} 
We now turn to multimode cavities, in which the interactions do not factorize as in $H_{\mathrm{1-mode}}$. The simplest case is the ring cavity, which supports two degenerate modes $\Xi_\pm(x) \sim e^{\pm ikx}$. In this case the interaction term takes the translation-invariant form

\beq\label{eq:ringcav}
H_{\mathrm{ring}} = - \zeta \sum_{i < j} \cos[k(x_i - x_j)] \, \mathbf{\sigma}_i \mathbf{\cdot \sigma}_j.
\eeq
Note that Eq.~(\ref{eq:ringcav}) is precisely Eq.~(1) of Ref.~\cite{weissman}, which approximately describes the RKKY interaction in materials (such as $\mathrm{Y}_x \mathrm{Gd}_{1-x}$) having spin susceptibilities peaked at a single momentum. While this interaction leads to frustration for Ising spins, it does not for XY spins; instead, the ground state is a spin spiral of pitch $\lambda$~\cite{ioffeJETP, weissman}. 

To realize frustration using XY spins, one must progress to cavity geometries possessing many degenerate modes, such as confocal and concentric cavities~\cite{siegman}. The general Hamiltonian for these is:

\beq\label{eq:mmgen}
H_{\mathrm{mm}} = - \zeta \sum_{\alpha, i \neq j} \Xi_\alpha(\mathbf{x}_i) \Xi_\alpha(\mathbf{x}_j) \mathbf{\sigma}_i \mathbf{\cdot \sigma}_j.
\eeq
For XY spins, Eq.~(\ref{eq:mmgen}) closely resembles the $O(2)$ generalizations~\cite{cook, nakamura:hopfield} of the Hopfield neural-network model~\cite{amit, hopfield}. The mapping to Cook's model~\cite{cook} is exact for translation-invariant, traveling-wave cavity geometries such as the ring and confocal cavities; however, the basic features 
of these models (which are similar to one another) are expected to extend quite generally to $H_{\mathrm{mm}}$~\cite{weissman}.

\textit{Associative memories, spin glasses, and self-organization.} 
The Hopfield and Cook models describe associative memories, consisting of $N$ neurons (i.e., spins in the physical system) that collectively encode $p$ ``patterns.'' In general, $p$ corresponds to the number of cavity modes that are resolvable, given the interaction range (i.e., $\chi$ in the notation of Refs.~\cite{us:pra}). The associative memory is said to function if, starting with any configuration similar to a stored pattern, the dynamics drives the configuration to the stored pattern, i.e., if a partially self-organized initial configuration at $T = 0$ becomes fully self-organized under the dynamics (this point is discussed further in Ref.~\cite{us:hopf}). In the Hopfield and Cook models~\cite{hopfield, cook, amit}, this is the case (as $N \rightarrow \infty$) for small $p/N (\alt 0.05)$, e.g., in the single-mode cavity. For $p/N \agt 0.05$, metastable states proliferate, and the system becomes a spin glass; the spin glass differs from the self-organized phase in that the ground-state atomic configuration \textit{does not} globally emit superradiantly into any particular cavity mode; nevertheless, it is a distinct phase from the high-temperature paramagnetic phase~\cite{binder}. A finite-temperature phase transition between the two is known to exist in the case that $p/N \rightarrow \infty$, i.e., the Sherrington-Kirkpatrick model~\cite{amit, binder}. These considerations lead to the global phase diagram shown in Fig.~\ref{fig:phasediag}. 

\textit{Tuning and detection}. Both the associative memory and the spin glass are low-temperature phases. The former is stable when $k_B T \alt \hbar \zeta N$. As is standard in ultracold atom experiments, the temperature is determined (in the $\kappa \ll \delta, \gamma \ll \Delta$ limit) by the system's initial entropy; however, $\zeta$ increases with pump laser intensity, and can be tuned across the transition. The spin ordering threshold is similar to that for self-organization, and is achievable, even for relatively large $\delta$, for reasonably long experimental lifetimes~\cite{baumann, us, us:pra}. The effective number of modes coupling to the atoms can be decreased by adjusting the length of the cavity away from the confocal/concentric limit; as $\zeta \sim 1/\delta$, only modes having sufficiently small $\delta$ couple strongly to the atoms. The spin glass transition temperature $T_g$ in Hopfield-type models is comparable to the single-mode ordering temperature~\cite{amit}. 

The self-organized phase should be detectable via the light emitted from the cavity, but the spin-glass phase is not, as it does not exhibit superradiance. One straightforward way to detect this phase is through its slow relaxational dynamics: a possible protocol involves initializing all spins in a certain region in the $\vert + \rangle$ state via a local spin addressing protocol~\cite{bloch2011}, and measuring the spin relaxation timescale (observed, e.g., via phase-contrast imaging~\cite{vengalattore}) as a function of pump intensity. A feature common to both the superradiant and spin-glass phases at low temperatures is the presence of a large number of low-energy excitations; these reveal themselves in condensed-matter systems via the heat capacity. In the cavity QED setting, such excitations can be detected, e.g., via two-photon spectroscopy~\cite{cornell:bragg}. Further possibilities for distinguishing the two low-temperature phases via their response functions are considered in Ref.~\cite{strack}.

\textit{Quantum regime}. Thus far, we have focused on the classical spin physics realizable using cavity-mediated interactions. We now turn to the quantum regime, in which Eq.~(\ref{eq:mmgen}) can be mapped~\cite{sachdev} onto a Bose-Hubbard model in the limit $U/t \rightarrow \infty$, via the transformation $\sigma_+ \rightarrow b^\dagger$:

\beq
H_{\mathrm{BH}} = -w \sum_{ij} t_{ij} (b^\dagger_i b_j + \mathrm{h.c.}) + \mu \sum_i b^\dagger_i b_i. 
\eeq
According to this mapping, a $\vert + \rangle$ state corresponds to the presence of a $b$ boson whereas a $\vert - \rangle$ state corresponds to the absence of a $b$ boson; the chemical potential $\mu$ is determined in the standard way from the number of bosons (i.e., $|+\rangle$ atoms). More generally, an $n$-state atom maps onto a Bose-Hubbard model with a maximum occupation per site of $n - 1$:

\beq
H_{\mathrm{n-level}} = -w \sum_{ij} t_{ij} (b^\dagger_i b_j + \mathrm{h.c.}) + U \sum_i (n_i - \bar{n})^2,
\eeq
where the ``interaction'' term $U$ can arise, e.g., because of the quadratic Zeeman shift. Note that, in this bosonic terminology, a ``superfluid'' state corresponds to a finite expectation value of $\sigma_\pm$ (i.e., to in-plane magnetic ordering). Thus, it is a ``superfluid'' of spins and \textit{not} of the atoms, which are frozen in place.

A crucial difference between cavity-based realizations of the Bose-Hubbard model and optical-lattice ones~\cite{mattP} is that, in the cavity-based setting, strongly disordered hopping amplitudes are natural (hopping amplitudes being determined by the oscillatory cavity mode functions) even in the \textit{absence} of on-site disorder (as both spin states interact identically with the trapping lasers). This is challenging to achieve in optical lattices~\cite{mattP}, as varying the hopping amplitude via the lattice depth inevitably leads to on-site disorder. The phase structure of the disordered Bose-Hubbard model is known to be richer in the absence of chemical potential disorder, especially in one dimension~\cite{altman:mottglass}: the off-diagonally disordered model exhibits a Mott glass phase, as well as a random-singlet glass phase~\cite{bhatt:lee, dfisher:singlets} (see Fig.~\ref{fig:phasediag}). These phases are not present in models having on-site disorder, and are thus not directly realizable in optical lattices, but are realizable in the cavity-based setting.

A simple one-dimensional geometry that realizes the model of Ref.~\cite{altman:mottglass} involves a chain of atoms trapped perpendicular to the cavity axis. The cavity modes are Hermite-Gaussian along this direction, which we label $y$; thus, the $n$th mode has a profile of the Hermite-Gaussian $H_n(y/L) \exp(-y^2/L^2)$, where $L$ is the waist of the TEM$_{00}$ mode. For good confocal cavities $n \sim 10-100$; a more scalable geometry involves atoms trapped along the cavity axis. An atom at position $y$ couples most strongly to modes with ``classical turning points'' near $y$; these modes have large amplitudes within a distance $L$ from the turning point, and either decay or oscillate rapidly beyond this distance. Thus the interaction range is given by $L$ and (in particular) is finite, and the analysis of Ref.~\cite{altman:mottglass} applies. For the case of a $\Lambda$ atom having equal spin populations, one can realize the random-singlet glass, which possesses long-range spin-singlet correlations~\cite{bhatt:lee, dfisher:singlets}. (In order to realize the Mott glass, one could use, e.g., an atom possessing three ground states.) The three realizable glassy phases can be distinguished, e.g., via internal-state dependent transport or compressibility measurements~\cite{mattP}.

\textit{Acknowledgments}. S.G. is indebted to Ehud Altman for helpful discussions. The authors acknowledge support from the DOE DE-FG02-07ER46453 (S.G.), NSF DMR 09-06780 (P.M.G.), and the David and Lucille Packard Foundation (B.L.L.).

\end{document}